\documentclass[11pt]{article}

\usepackage{graphicx}
\usepackage{array}

\newcommand{\bm}{\boldmath}

\title{{\bf Study of atomic motions in EuBa$_2$Cu$_3$O$_{7-\delta}$ using M\"ossbauer
and EXAFS spectroscopies} \\
{\small {\bf Short title}: atomic motions in EuBa$_2$Cu$_3$O$_{7-\delta}$}}

\date{}

\begin{document}

\maketitle

\author{F. Piazza$^1$, E. Abraham$^1$,  L. Cianchi$^2$, F. Del Giallo$^2$, P. Ghigna$^3$, 
        F. Allegretti$^4$ and G. Spina$^4$}

\begin{itemize}

        \item[$^1$] {\small \em Heriot--Watt University, Physics Dept - EH14 4AS, Edinburgh, U.K.} 
        \item[$^2$] {\small \em IROE, CNR - via Panciatichi 64, 50127 Firenze, Italy} 
        \item[$^3$] {\small \em Universit\`a di Pavia, Dip. Chimica Fisica - v.le Taramelli 16, 27100
                     Pavia, Italy} 
        \item[$^4$] {\small \em Universit\`a di Firenze, Dip. Fisica - via di S. Marta 3, 
                      50139Firenze, and INFM}

\end{itemize}

\bigskip
{\bf Pacs} 74.72.B

\begin{abstract}

We report temperature--dependent M\"ossbauer and EXAFS measurements of atomic mean square
displacements on different samples of
EuBa$_2$Cu$_3$O$_{7-\delta}$. Our results indicate that the collective atomic motions
are characterised by large--amplitude anharmonic oscillations.

\end{abstract}
\newpage
\section{Introduction}

Numerous studies of the vibrational features of superconducting cuprates
have shown that some ions in the unit cell of these materials perform
large--amplitude anharmonic oscillations even at low temperatures (see
{\em e.g.} \cite{Capaccioli} and references therein). However, in some cases the results
obtained by means of different experimental techniques are not in agreement. For example,
both neutron--diffraction
\cite{Schweiss} and EXAFS experiments \cite{Mustredeleon} reveal vibrational
anharmonicity in the motion of apical oxygens. However, neutron diffraction on
samples of YBCO single crystals ascribes the anharmonicity to motions in the
($\hat{a,b}$) plane, while EXAFS ascribes it to the motion along
$\hat c$. Another example: neutron diffraction performed at 10 K on
samples of RBCO powders (R= Y and rare earths) \cite{Guillaume}
showed that the R-ion motion is practically independent of the ion's
atomic number, and is consistent with normal
atomic vibrations. This result has also been confirmed by neutron
diffraction on samples of RBCO single crystals (R=Y,Ho)\cite{Schweiss}. However,
M\"ossbauer spectroscopy performed on samples of EuBCO powders measures a
much larger mean square displacement (MSD) for the Eu ion \cite{Capaccioli}. Moreover,
the M\"ossbauer data on samples of EuBCO oriented powders, which will be presented
in this paper, also confirm the large low--temperature anharmonicity of the Eu ion,
and show that this anharmonicity must be ascribed to the
component along $\hat c$. 
Since there is no plausible reason for the Eu ion to behave differently from Y
and the other rare earth ions, we must conclude that M\"ossbauer data
conflict with the neutron ones. 

As is well known, the pairing mechanism in high-$T_c$
superconductors is not yet understood. In particular, the role played by the
lattice thermal vibrations is still debated
(see {\em e.g}~\cite{lattice} and references therein). Hence, it is very important to
deeply  investigate such low-temperature anharmonicities. In
order to prove their existence, the MSD must be
determined as a function of the temperature. For  harmonic crystals, 
the high temperature trend of the atomic MSD extrapolates through the
origin \cite{Kolk}. On the contrary, if the ion moves in a non--parabolic potential, the
extrapolation of the high--temperature trend intercepts the y--axis above the origin.
The $T=0$ intercept provides a measure of the ``flat'' part in the potential--well
bottom. 

In this article, we present a study of the vibrational features of EuBCO as seen
by using both M\"ossbauer and EXAFS spectroscopies. In particular, 
M\"ossbauer spectra of $^{151}$Eu in samples of oriented powders will be analysed, in
order to investigate the anharmonic components of the motion. On the other hand, EXAFS
measurements at the  Eu K-edge make it possible to obtain the mean square relative
displacements (MSRD) of the Eu ion  and its neighbours. In summary, information on 
the correlation among atomic motions in the unit cell can be
obtained by taking into account M\"ossbauer and EXAFS data.

\section{M\"ossbauer data}

Two samples of oriented single--phase powder (samples $1$ and $2$) and a sample of 
randomly-oriented single--phase powder (sample $3$) of EuBa$_2$Cu$_3$O$_7$  were prepared, all in
the form of flat disks. The orientation was obtained by means of two different methods: (1) we
mixed the finely--ground powder  with epoxy resin, and shaped it in a  8 T magnetic field at room
temperature. By doing this, for EuBCO, as for ErBCO, the monocrystalline grains oriented
themselves with the $\hat c$ axis perpendicular to the magnetic-field \cite{Knorr, Durand}.
(2) We shaped the powder -- a mixture of EuBa$_2$Cu$_3$O$_7$ and polyethylene -- 
under  a pressure of about 0.3 GPa, thereby orienting the grains with the $\hat c$ axis
in the direction of the pressure. Finally,  we obtained sample 3
simply by mixing the powder with epoxy resin. The
orientation degree was determined by means of X--ray Bragg diffraction with 
Bragg--Brentano acquisition geometry. The orientation distribution is Gaussian in the cosine
of the orientation angle for
both samples, with standard deviations of 15$^0$ for the magnetically oriented
sample and about twice as much for the one oriented under pressure. 
The transition temperature for all the samples is 92 K.

\begin{figure}[h!]
\begin{center}
\includegraphics[height=6 truecm]{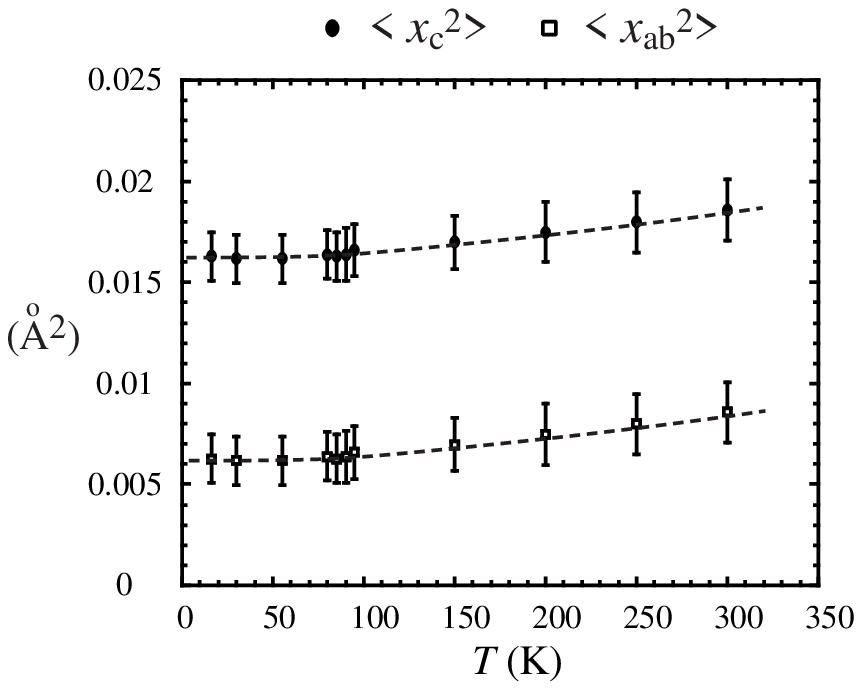}
\caption{Trend of the MSD along $\hat c$ and in the $\hat {ab}$ planes  as a function
of the temperature}
\label{f.1}
\end{center}
\end{figure}

We collected six spectra at the temperatures T=20, 85, 90,
95, 200 and 300 K  for each sample, using a standard spectrometer with a SmF$_3$ source
of 3.7 GBq at room temperature. From the analysis of the spectra, 
we calculated the ratio between the Debye-Waller factors in
the $\hat c$ and $(\hat {a,b})$ directions as functions of the temperature.
We carried out this analysis for the two oriented samples, combining them with the 
results from the isotropic sample. We
got the same value for the ratio $f_c/f_{ab}$ in the two cases, thereby
obtaining a good cross--check for the reliability of the measurement. 
This also provided direct evidence of consistency of data sets from the different
grain--aligned samples. A detailed description of this analysis will be given
elsewhere\footnote{work in preparation}. The
$f_c/f_{a,b}$  ratio turned out about 0.4 at all temperatures. This means that the
motion of the Eu ion is anisotropic, the amplitude in $\hat c$ direction being greater
than the one in the 
$(\hat {a,b})$ plane. By combining this result with the previous  M\"ossbauer measurements
 of the  angular--averaged $^{151}$Eu MSD \cite{Capaccioli}, we can calculate 
$\langle x_c^2\rangle$ and $\langle x_{ab}^2\rangle$. The results of these calculations
are plotted in fig.~\ref{f.1}. We note that the MSD of the $\hat c-$component is about
three times greater than the $(\hat a,\hat b)-$component. Moreover, the $T=0$ intercept
of the high temperature trends of both components do not cross the origin.
Consequently, our M\"ossbauer data show low-temperature anharmonicities, both along $\hat
c$ and in the $(\hat a, \hat b)$ plane.

\section{EXAFS data}

 We prepared two samples of EuBa$_2$Cu$_3$O$_{7-\delta}$ single--phase powders  mixed
with polyethylene, with $\delta\approx1$ (nonsuperconducting)  and $\delta\approx0$
($T_c \approx 90$ K). For each
of them, we collected trasmission Eu K--edge EXAFS spectra at $T=$30, 60,
90, 120,180, 250 and 300 K. The measurements were performed at the beam--line GILDA CRG
of the European Synchrotron Radiation Facility (ESRF) in Grenoble. 

\begin{figure}[h!]s
\begin{center}
\includegraphics{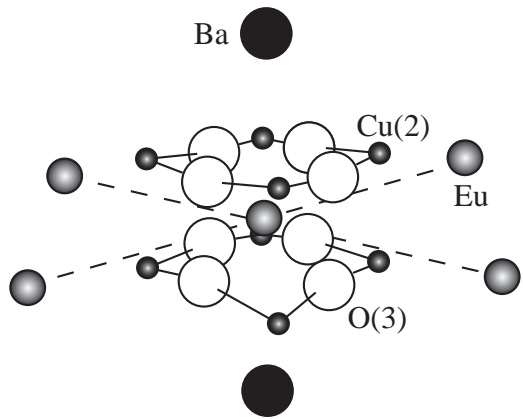}
\caption{Cluster used in the simulation of the EXAFS spectra. In order of increasing
distance from the Eu site, one can see the O shell, the Cu shell, the Ba shell and the
shell of  Eu's lying in the adjacent cells.}
\label{f.2}
\end{center}
\end{figure}

\begin{figure}
\begin{center}
\includegraphics{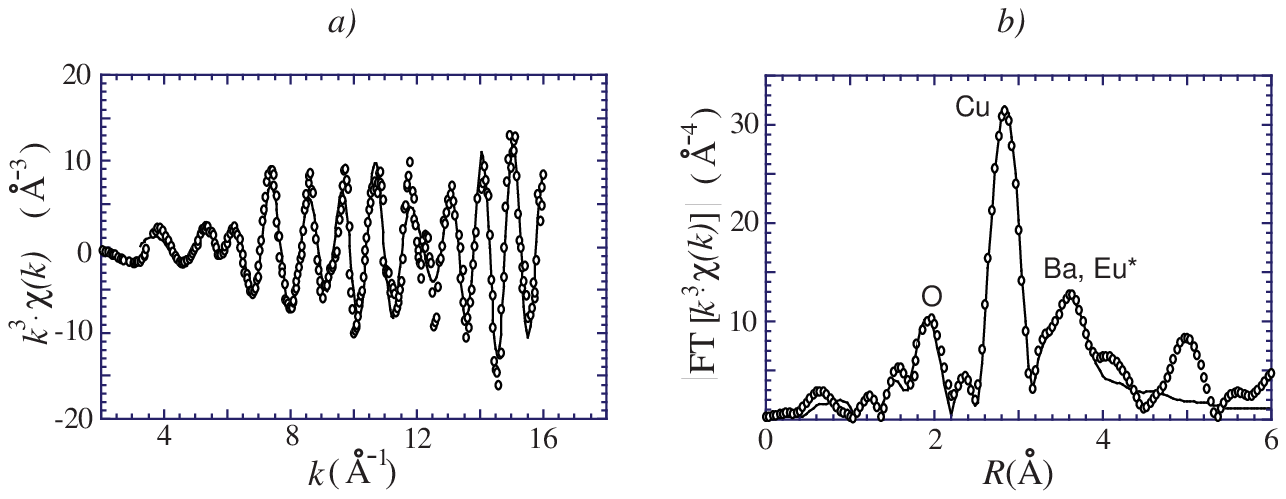}
\caption{{\em a)} Eu K-edge EXAFS signal measured in the underdoped sample at a
representative temperature $T = 30$ K multiplied by $k^3$. Circles are experimental
points. The solid line is the result of the fit. {\em b)} Magnitude of the Fourier
transform of the representative signal shown beside. Small circles: experimental. Solid
line: fit. The fit of the signal for $k\approx12$ \AA$^{-1}$ is less accurate. This is
because the contribution to the spectra for $k\approx12$ comes from the
remote shells, which are not taken into account in our signal calculation. }
\label{f.3}
\end{center}
\end{figure}

The model cluster used for the EXAFS calculations is shown in fig.~\ref{f.2}, which
represents the local structure around Eu in EuBCO up to a radial distance of 4\AA. The
four shells included in the calculations are denoted as Eu--O, Eu--Cu, Eu--Ba and
Eu--Eu(2), in order of (their) increasing distance from Eu. Actually, the EXAFS Fourier
transform shows further components above 4 \AA, but their analysis is much less accurate.
Therefore, they have not been considered in fitting the experimental data. A typical EXAFS
signal with corresponding Fourier Transform is shown in fig.~\ref{f.3}.

We carried out the data analysis by fitting the experimental spectra with the aid of
the GNXAS package \cite{gnxas}, with distances $R_i$ and Debye-Waller factors $\sigma^2_i$
as fitting parameters, beside the energy treshold $E_0$ used in the energy--to--wavevector
conversion.  We found that the average distances between europium and its
neighbours do not vary appreciably (less than 0.5\% for temperatures up to 300 K). The
trends of the MSRD's are more interesting. The best fit results  are reported in
table~\ref{t.1} and table~\ref{t.2} for the samples with $\delta \approx 0$ and $\delta
\approx 1$, respectively. It can be seen that they lie in a range which is consistent with
normal harmonic vibration (see {\em e.g.}
\cite{Greegor}, where the $\sigma^2$ values for copper are reported). Moreover,
the $\sigma^2$'s of the first two shells do not depend on the doping. On the contrary, the
$\sigma_i^2$'s corresponding to the Ba and Eu(2) shells are observed to be
somewhat smaller for the overdoped sample.

\section{Correlations among ionic motions} 

As is well known, the EXAFS Debye--Waller factors measure the broadening of
the bond distances due to the thermal motions, provided the static disorder
is negligible. Then, denoting by $\sigma_1^2$ and
$\sigma_2^2$ the absorber and the backscatterer MSD's along the interatomic 
distance $\bm R_{12}$  respectively, the corresponding
$\sigma_{EXAFS}^2$ is given by:

\begin{equation}
\label{e.1}
\sigma_{EXAFS}^2=\sigma_1^2+\sigma_2^2-2\rho_{12}\sigma_1\sigma_2
\end{equation}
where $\rho_{12}$ denotes the so--called correlation parameter. If $\rho_{12}\approx
\pm1$, the motion is highly correlated, in--phase or out--of--phase, respectively. If
$\rho_{12}\approx 0$, the ionic oscillations are uncorrelated. Generally, a decrease
of
$|\rho_{12}|$ is expected as the interatomic distance increases. Furthermore, if
the Debye model is a good approximation for the considered solid, $\rho_{12}$ is
expected to be  an increasing function of temperature. This is a consequence
of the correlations, as seen by EXAFS, being projected onto the 
absorber--backscatterer bond distance, thereby being 1D correlations
in their very nature~\cite{Beni}. 

Some qualitative conclusions regarding the correlations among  ion motions
are drawn here as follows. Let us  consider the ionic radii of Eu$^{3+}$ and
O$^{2-}$. We see that the two ions are practically in contact with each other~\cite{ions}.
Consequently, a large positive correlation between the motions of the europium and the
oxygen ions of the first shell is expected. 
Moreover, taking into account the tight binding between the oxygen and copper ions in
the CuO$_2$ planes, a large positive correlation is  also expected with the copper ions of
the second shell  \cite{Lanzara}. On the other hand, we note that Ba$^{2+}$ ions are
not in contact with the inner shells. Nevertheless, the corresponding measured $\sigma^2$
show to be rather small. Then, in order for these results to be consistent with the large
oscillation of the Eu ion along $\hat c$ as measured by M\"ossbauer spectroscopy,  a
large positive correlation between the two ions has to exist. Finally, we can combine
the M\"ossbauer value for the $\sigma_{\rm Eu}^2$ in the $(\hat a, \hat b)$ plane with
the $\sigma^2_{EXAFS}$ to calculate $\rho_{\rm Eu-Eu}$ as a function of temperatuure.
Unlike the normal trend, $\rho_{\rm Eu-Eu}$  decreases monotonically from $\approx
1$ to $\approx0.75$ as the temperature increases from 30 to 300 K.

\section{Conclusions}

\begin{figure}
\begin{center}
\includegraphics{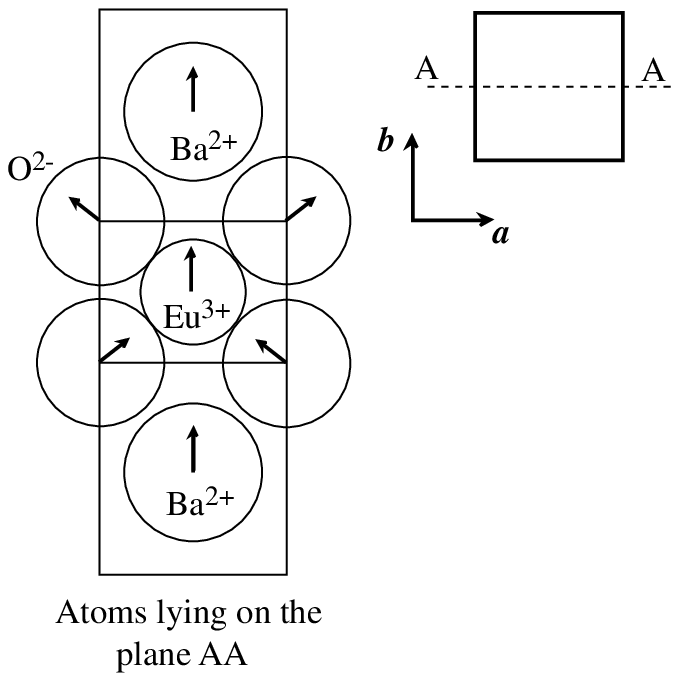}
\caption{Schematic representation of the O and Ba shells in an EBCO unit cell drawn to
scale. The arrows represent large displacements of the ions, which are caused by a large
displacement of the europium ion along $\hat c$.}
\label{f.4}
\end{center}
\end{figure}

By combining the M\"ossbauer and EXAFS results  with
the expected scenario of a large positive correlation between the motions of Eu and O ions,
we suggest that the ionic motions consist of large--amplitude, highly--correlated
oscillations with superimposed small--amplitude, weakly--correlated, vibrations.
The large--amplitude component can be described as follows. Starting from
the equilibrium configuration, let us imagine we displace the Eu ion along $\hat c$.
As a consequence, the oxygen ions will move along the line joining the Eu and O
centres, see fig.~\ref{f.4}. But this displacement hardly changes the Eu--O distance and
therefore, it will not be revealed by EXAFS. Hence, the measured
$\sigma^2_{\rm Eu-O}$'s  correspond only to the
small--amplitude component. The same considerations apply to the Cu ions, since their
motions are closely correlated with the motions of the oxygen neighbours \cite{Lanzara}. 

The rather small values of $\sigma^2_{\rm Eu-Ba}$ measured in our experiment may
seem surprising. In fact, since Ba$^{+2}$ ions are not in contact with the inner shells,
the large--amplitude oscillations of Eu along $\hat c$, as measured by M\"ossbauer
spectroscopy, would result in a large--amplitude relative motion for these two ions.
However, $\sigma^2_{\rm Eu-Ba}$'s are fairly small. Consequently, we should conclude that 
the motions of barium and europium are strongly and positively correlated, despite their
rather large separation. A possible explanation could be that, due to coulomb interaction
between the Eu$^{+3}$ and Ba$^{+2}$ ions, when the former moves in a certain direction the
latter is forced to move in the same direction, and {\em viceversa}. We also note that the
large displacements of the oxygen ions result in a variation of the sum of their attractive
forces on the barium ions. This, in turn, increases the correlation between europium and
barium motions. To summarise, the large anharmonic oscillation
includes the Eu  and Ba motions along $\hat c$ and the rotation around the cell
centre of the oxygen ions of the first shell. 

Finally, we observe that our experiments do not reveal remarkable differences between
the underdoped and overdoped samples.  The $\sigma^2_{\rm Eu-O}$ and
$\sigma^2_{\rm Eu-Cu}$ values clearly show that the relative motions of these ions are
independent of the doping level. On the contrary, the  $\sigma^2_{\rm Eu-Ba}$
and $\sigma^2_{\rm Eu-Eu}$ values of the overdoped sample are significantly smaller than
those of the underdoped one at $T<T_c$. Such a difference could be due to the coupling
phenomenon that takes place at $T<T_c$ in the superconducting sample. However, in
order to clarify this question, further investigations will be necessary. We also note that
the observed decrease in $\rho_{\rm Eu-Eu}$ as the temperature increases cannot be
explained within the framework of phonon theory.


\newpage
\centerline{\bf \Large Figures and Tables}
\vspace{1 truecm}

\newpage

\begin{table}[h!]
\caption{EXAFS-MSRD's for the underdoped EBCO.}
\label{t.1}
\begin{center}
\begin{tabular}{ccccc} \hline
T (K) \vspace{0.2truecm} & $\sigma^2_{\rm O}$ (10$^{-3}$\AA$^2$) & $\sigma^2_{\rm Cu}$ (10$^{-3}$\AA$^2$)
&$\sigma^2_{\rm Ba}$ (10$^{-3}$\AA$^2$) &$\sigma^2_{\rm Eu}$ (10$^{-3}$\AA$^2$) \\ 
\hline
30  & 4.0(5) & 2.1(2) &  3.0(6) & 2.2(3) \\
60 & 4.1(5) & 2.2(2) &  3.5(7) & 2.2(3) \\
90  & 4.4(5) & 2.5(2) &  3.9(8) &2.4(3) \\
120  & 4.7(6) & 2.9(2) &  4.1(9) & 3.2(4) \\
180 &5.5(7)& 3.7(3) & 4.6(10) & 4.2(5) \\
250 &6.7(9) & 4.7(3) & 5.5(16) & 5.1(8) \\
300 \vspace{0.2truecm} &8.1(11)& 5.2(4) &  7.4(19) & 6.0(10) \\ \hline
\end{tabular}
\end{center}
\end{table}

\begin{table}
\caption{EXAFS-MSRD's for the overdoped EBCO.}
\label{t.2}
\begin{center}
\begin{tabular}{ccccc} \hline
T (K) \vspace{0.2truecm} & $\sigma^2_{\rm O}$ (10$^{-3}$\AA$^2$) & $\sigma^2_{\rm Cu}$ (10$^{-3}$\AA$^2$)
&$\sigma^2_{\rm Ba}$ (10$^{-3}$\AA$^2$) &$\sigma^2_{\rm Eu}$ (10$^{-3}$\AA$^2$) \\ \hline
30  & 4.2(7) & 2.0(2) &  1.4(4) & 1.3(3) \\
60 & 4.2(7) & 2.0(2) &  1.4(4) & 1.3(3) \\
90  & 4.4(8) & 2.5(2) &  2.4(6) &1.6(3) \\
120  & 4.7(8) & 2.8(2) &  2.7(7) & 2.3(8) \\
180 &5.3(10)& 3.4(3) & 2.8(9) & 3.2(12) \\
250 &6.8(12) & 4.8(3) & 4.4(14) &5.3(18) \\
300 \vspace{0.2truecm} &7.5(12)& 5.2(4) &  4.8(12) & 5.7(16) \\ \hline
\end{tabular}
\end{center}
\end{table}


\end{document}